\documentstyle[emulateapj]{article} 
\slugcomment{submitted to ApJ part two} 
\date{Jan.14, 1999} 
\lefthead{} 
\righthead{} 
 
\begin{document} 
\title{Is Coherence Essential to Account for Pulsar Radio Emission?} 
\author{Bing Zhang\altaffilmark{1}, B. H. Hong \altaffilmark{2},  
    G. J. Qiao\altaffilmark{3}} 
\affil {Chinese Academy of Science - Peking University joint Beijing 
 Astrophysics Center (BAC), \\
 and Department of Astronomy, Peking University, 
 Beijing 100871, P.R. China}
\altaffiltext{1} {NRC Research Associate Fellow,
 Laboratory of High Energy Astrophysics, Code 661, NASA Goddard Space
 Flight Center, Greenbelt, MD 20771, USA. Email: 
 bzhang@twinkie.gsfc.nasa.gov, zb@bac.pku.edu.cn}
\altaffiltext{2} {Email: hbh@bac.pku.edu.cn} 
\altaffiltext{3} {CCAST (World Laboratory) P.O.Box 8730, Beijing 
 100080, P.R.China. Email: gjn@pku.edu.cn} 

\date{2-Dec-1998} 
 
\begin{abstract} 
 Based on definitions, two joint-criteria, namely, the optical-thin 
 constraint and the energy budget constraint, are proposed to judge whether 
 the emission nature of radio pulsars is incoherent or obligatory to be
 coherent. We find that the widely accepted criterion, $kT_{\rm B}\le 
 \epsilon$, is not a rational criterion to describe the optical-thin 
 condition, even for the simplest case. The energy budget constraint 
 could be released by introducing a certain efficient radiation mechanism
 (e.g. the inverse Compton scattering, QL98) with emission power
 of a single particle as high as a critical value $P_{\rm sing,c}\sim
 10^{-4}-10^{-3} {\rm erg \cdot s}^{-1}$. This in principle poses
 the possibilities to interpret high luminosities of pulsars in terms
 of incoherent emission mechanisms, if the optical-thin constraint could
 be released by certain mechanism as well. Coherence may not be an
 essential condition to account for pulsar radio emission.

\end{abstract} 
 
\keywords{pulsar: general - radiation mechanism: non-thermal} 
 
\section{Introduction} 
Shortly after the discovery of the first pulsars, the question why they  
pulsate is soon answered in terms of rapid spin of the magnetized neutron  
stars. Another fundamental question why they shine, however, is not fully  
answered till today, after 30 years. The key problem lies in the observed 
high brightness temperatures of pulsar emissions, which are commonly thought 
to be due to certain coherent radiation mechanisms (Ginzburg, Zheleznyakov 
\& Zaitzev, 1969, hereafter GZZ69; Ginzburg \& Zheleznyakov 1975; 
Melrose 1992a,b, hereafter M92a,b). Various coherent mechanisms have been 
proposed to interpret the nature of pulsar emission (for reviews, see e.g. 
Ginzburg \& Zheleznyakov 1975; M92a,b).
 
Here we'll re-investigate the fundamental question whether coherence is
an obligatory condition for any promising emission model of radio pulsars.
We'll show that the widely adopted criterion, i.e. 
$$kT_{\rm B}\leq \epsilon, \eqno(1)$$
where $T_{\rm B}$ is the brightness temperature of radiation, $\epsilon=
(\gamma-1)mc^2$ is the kinetic energy of the relativistic particles, is 
not a rational criterion to judge whether a radiation mechanism is
obligatory to be ``coherent''. We'll put forward two joint-criteria by
definitions instead and show how severe these constraints are on any 
incoherent mechanism. 

\section{What's the key criterion to judge the coherent nature of  
pulsar emission?} 

\subsection{Incoherent and coherent emission}
According to M92a, there is no general definition of the so-called
``coherent'' emission, while a working definition is ``any non-thermal 
emission that cannot be explained in terms of incoherent emission''. For
an incoherent mechanism, it is defined as that ``the intensity (or luminosity) 
of radiation from the source is less or equal to the sum of the intensities 
of spontaneous emission from different parts of the source'' (GZZ69), and 
as ``due to spontaneous emission when the corresponding absorption process 
is unimportant'' (M92a). From these definitions, we come to two
inequaties as the constraints of any incoherent emission, which read
$$\tau=\int_0^{s_{\rm e}}\alpha_\nu {\rm d}s^\prime \le 1,  \eqno(2a)$$
and
$$P_{\rm sing} n l^2 s \ge L_{\rm obs}. \eqno(2b)$$
The constraint [2a] is to define that ``absorption process is unimportant'', 
in which $\alpha_\nu$ is the absorption coefficient, $\tau$ is the 
optical-depth of the radiation produced in the inner magnetosphere which
travels an escaping distance $s_{\rm e}$ to out of the magnetosphere. We also
define $s_0$ as the upper limit of $s_{\rm e}$ defined by $\tau=1$, which is 
the ``transparent'' distance of the radiation. We call [2a] the ``optical-thin 
constraint''. The next constraint [2b] is to show that the sum of the 
emission power from individual particles (i.e. $P_{\rm sing}$) could meet 
the total emission power (i.e. luminosity $L_{\rm obs}$) observed, which we 
call the ``energy budget constraint''. Here $n$ is the number density of the
particles who contribute to the luminosity, and the volume is adopted as 
$l^2 s$, where $l$ is the typical extended scale of the source and $s$ is
the depth of the emission region. Note $s\leq s_0$, since emission deeper
that the ``transparent'' depth $s_0$ can not escape from the magnetosphere 
(see [2a]). Any violation of the either constraint above will make an 
incoherent emission failed to interpret the observed emission from pulsars.

Generally, people tend to think that the high brightness temperatures 
inferred from pulsar observations have made both constraints failed, which
means that one has to appeal to certain coherent mechanisms. The earliest 
idea of coherent mechanism is to solve the ``energy budget'' problem by 
introducing the ``bunching'' mechanism (e.g. Ruderman \& Sutherland 1975). 
This will enhance the emission power of an individual particle to a factor 
of $N$, where $N$ is the number of the particles in the bunch. Such a mechanism 
was later criticized both about its origin and its maintenance (Melrose 1978; 
hereafter M78; M92a,b). Furthermore, it was argued that such a coherent 
mechanism still fails to fulfill the ``optical-thin'' condition [2a] provided 
the typical brightness temperatures inferred from pulsars (M78). To release 
the optical-thin constraint, usually certain ``maser'' (negative absorption)
mechanisms are proposed. In the maser models, the optical depth becomes
negative so that [2a] is well satisfied, and the ``amplification'' effect
due to population reverse of the energy levels can meet the energy budget 
requirement. The problems in these models, however, lie in that there is no 
simple form of curvature maser mechanism to operate, and that more complicated 
forms are apparently in contradiction to certain observational or ``inferred'' 
facts. For example, maser curvature emission including curvature drift (Luo \& 
Melrose 1992) rules out the possible emission from the millisecond pulsars, 
while the fact is that there is no distinct difference between the emission
features of millisecond pulsars and those of normal pulsars (Manchester 1990). 
The third group of the coherent mechanisms are the even more complicated forms 
of ``indirect'' emissions by invoking ``hydrodynamic instabilities'' (e.g. 
Beskin, Gurevich, \& Istomin 1988; Kazbegi, Machabeli, \& Melikidze 1991), 
some forms of which are regarded as the ``most plausible'' mechanisms (M92b).

\subsection{Is $kT_{\rm B} \le \epsilon$ a rational criterion to judge
the coherent nature of the source?}
The point which drives people thinking that coherence is essential is the
widely adopted criterion [1] (e.g. GZZ69; M92a, b; Luo 1998) rather than 
criteria [2a,b]. According to [1], to achieve the observed high brightness 
temperatures of pulsars (typically $10^{27}$K at 400 MHz, Sutherland 1979;
also see eq.[3] below) with an incoherent process, 
the Lorentz factor of the particles should be as high as $\gamma-1\ge 1.7
\times10^{17}T_{\rm B,27}$, which is unachievable in various pulsar polar 
cap models due to the limitation of the inner accelerators by the $\gamma-B$ 
(e.g. Sturrock 1971; Ruderman \& Sutherland 1975; Arons \& Scharlemann 1979; 
Usov \& Melrose 1996; Zhang et al. 1997a; Harding \& Muslimov 1998) or 
$\gamma-\gamma$ (Zhang \& Qiao 1998) absorption of the $\gamma$-rays. 
In other words, for a certain $\gamma$ of the particles, the brightness 
temperature is constrained to $T_{\rm B}\leq 5.9\times 10^9$K $\gamma$. 
This leads to the conclusion that any emission model {\em must} appeal to 
certain coherent mechanisms. Here we'll re-investigate the validity of 
eq.[1] and show that it is not a rational criterion.

The so-called brightness temperature $T_{\rm B}$ of a certain object is 
defined as the effective temperature of a blackbody whose emission  
intensity at a certain frequency is just equal to the observing intensity 
of the object at that frequency. In the Rayleigh-Jeans regime which is  
viable for pulsar radio emissions, the brightness temperature reads 
$T_{\rm B}=(c^2 / 2\nu^2 k) I_\nu$ (Rybicki \& Lightman 1979), where 
$k$ is the Boltzmann's constant, $I_\nu$ is the radiation intensity at 
the frequency $\nu$. Connecting $I_\nu$ with direct observables, i.e. 
the flux $F_\nu=I_\nu \cdot\Delta\Omega_s$ (in unit of mJy) and distance $d$  
(in unit of kpc), where $\Omega_s=l^2 / d^2$ is the solid angle the source
opens to the telescope, and again $l$ is the extended scale of the source 
region with typical value of $10^6$cm (Manchester \& Taylor 1977; Lesch
et al. 1998), we come to
$$T_{\rm B}\simeq 3.1\times 10^{23}{\rm K}\nu_9^{-2}  
F_\nu({\rm mJy})d^2({\rm kpc}) l_6^{-2}, \eqno(3)$$ 
where $l_6$ is $l$ in unit of $10^6$cm, and $\nu_9$ is $\nu$ in unit of 
GHz. Note this temperature is different from the antenna temperature of 
the telescope when $\Delta\Omega_s$ is smaller than the beam solid angle 
of the telescope $\Delta\Omega_A$, which is usually the case for pulsars. 
Thus eq.[3] is a pure theoretical value irrelevant to the telescope details.
We see this temperature is very high, and even higher for lower frequencies
(e.g. 400 MHz).

The criterion [1] directly comes from the study of synchrotron 
self-absorption sources. However, it was not proved in the case of 
pulsars to our best knowledge. According to M92a, eq.[1] is the result 
of the limitation of self-absorption, thus it might be a variant of 
condition [2a]. Under the {\em isotropic} condition, the absorption 
coefficient can be deduced as
$$\alpha_\nu={(p+2)c^2 \over 8\pi\nu^2}\int {\rm d}\epsilon P(\nu,\epsilon)
{N(\epsilon)\over\epsilon} \eqno(4)$$
under the condition of $h\nu \ll\epsilon$, where energy-dependent particle 
density $N(\epsilon)$ has a power law distribution with index $p$, and 
$P(\nu, \epsilon)$ is the spectral emission power of the particle 
with energy $\epsilon$ (Rybicki \& Lightman 1979, their eq.[6.52]). 
Adopting a very simple form of $p=1$
and $N(\epsilon)=n/\epsilon$, and submitting them to eq.[2a], we get a 
dimensional analysis estimate limit as $(3/ 8\pi)(c^2/\nu^2) P_\nu n s
\le\epsilon$. Note the luminosity $L_{\Delta\nu}=(4\pi f)F_\nu d^2\Delta\nu
=P_\nu nl^2 s\Delta\nu$, where $(4\pi f)$ is the beaming factor of the
emission ($f=1$ for isotropic emission), again using $T_B=(c^2/2\nu^2 k)
I_\nu$, the inequality [2a] finally turns out to be 
$$kT_{\rm B}\le (\epsilon/3f). \eqno(1')$$ 
Note $[1']$ is identical with [1] only when $f\sim 1$, i.e., for isotropic
radiation. This could be the case of a synchrotron self-absorbed source,
but never true for the case of pulsars which has $f\ll 1$. This means that
even adopting an isotropic absorption formula as [4], and for the simplest 
case, eq.[1] is not a rational criterion at all.

\subsection{Optical-thick or optical-thin?}

Though $[1']$ sets a less stringent constraint to the achievable brightness
temperature than [1] does, it does not release the optical-thin constraint, 
since high beaming effect (small $f$ or small $l$) also raises the inferred
value of $T_{\rm B}$ (see eq.[3]). However, we suppose that when we
present explicit studies on more detailed mechanisms, the constraint
[2a] using [4] could come to much different results than $[1']$.
Furthermore, even eq.[4] may not be true when considering the complicated
situation (e.g. anisotropic and far-from-equilibrium) in pulsar magnetospheres 
(see below). Thus in judging whether coherence is obligatory for a certain 
mechanism, we have to always go back to criteria [2a,b]. Neither [1] nor 
$[1']$ make much sense.

From the observational point of view, it seems like that pulsar 
magnetospheres might be optical-thin. Clear beam ``radius-to-frequency
mapping'' has long been observed from the multi-frequency observations 
(e.g. Rankin 1983), which means that we have observed different ``depth'' 
in the pulsar magnetospheres. It is hardly believable that most pulsar 
emissions are buried inside the optical-thick layers, while only a small 
portion on the surface layer escapes.

One can argue that such a optical-thin feature could be the result of
certain maser (or other forms of) coherent mechanisms. However, 
we see that constraint [2a], which requires $\tau \le 1$, still 
leaves much room to solve the problem
without invoking any maser mechanisms which requires $\tau < 0$.
Though proposing such a detailed optical-thin incoherent theory 
is beyond the scope of this paper, we present here some criticisms
on the present studies on the absorption processes in pulsar magnetospheres.
These comments are quite preliminary, but nevertheless some possibilities
to solve the optical-thin problem within the framework of ``incoherent''
emission mechanisms. We hope these ideas could promote more profound 
studies in this area. First, almost all the studies on the self-absorption
processes in pulsar magnetospheres have inherited the formalism derived 
from the stellar atmospheres, which is in {\em isotropic} condition (see [4]). 
The intense magnetic fields in pulsar magnetospheres, however, have destroyed 
such isotropy so that the particle motion is essentially one-dimensional as
the particle is bound to the lowest Landau state. As a result, the photon 
field also have a preferred emission direction along the field line and 
particles are usually interacting with the photons tail-on. Thus absorption 
should be angle-dependent, which might lead to quite different results than 
eq.[4]. Another point is that the situation in
pulsar magnetosphere is {\em far from equilibrium}, while the formalism
people tend to use are derived from near-equilibrium regime. Some possible
non-linear effect may be able to solve the optical-thin problem. 

%More basically, the Einstein relations, which are argued to be an intrinsic 
%properties of the atoms, are never proven strictly in the case of pulsar,
%where only charge-separated electron-positron plasma exists and both
%plasma and photon fields are far from equilibrium.

\subsection{Energy budget} 
 
Suppose constraint [2a] could be released by introducing certain 
incoherent mechanism, is coherent still obligatory by constraint [2b]? 
Let's express [2b] in a more obvious way.

The luminosity $L_{\Delta\nu}$ in a certain frequency band $\Delta\nu$ 
can be expressed in direct observables by
$$L_{\Delta\nu}\simeq 9.5\times 10^{25}{\rm erg\cdot s}^{-1}(4\pi f) 
F_\nu({\rm mJy})d^2({\rm kpc})\Delta\nu_9. \eqno(5)$$ 
This defines a critical value of the emission power of a single
particle $P_{\rm sing,c}=(L_{\Delta\nu}/ n l^2 s)$. Assuming the particle
number density is $n=\xi n_{\rm gj}$, where $n_{\rm gj}\simeq(\Omega
B/2\pi ce)\simeq 7.0\times 10^{10}{\rm cm}^{-3}B_{12}P^{-1}$ is the
Goldreich-Julian density of pulsar magnetosphere (Goldreich-Julian
1969) and $\xi$ is the multiplicative factor due to the pair cascades, 
this critical power turns out to be
$$P_{\rm sing,c} \simeq 1.4
\times 10^{-4}{\rm erg\cdot s}^{-1}(4\pi f) F_\nu({\rm mJy})
d^2({\rm kpc})$$
$$ \times \Delta\nu_9 P_{0.1}B_{12}^{-1}\xi^{-1}l_6^{-2}s_6^{-1}.
\eqno(6)$$ 
Any mechanism with $P_{\rm sing} >P_{\rm sing,c}$ could meet the energy
budget constraint. Note $l=10^6$cm$l_6$ and $s=10^6$cm$s_6$ are
adopted as the typical linear size of the emission region which 
contributes to the pulsar luminosity (Manchester \& Taylor 1977; Lesch
et al. 1998). This is a little bit arbitrary, and $P_{\rm sing,c}$ is 
quite sensitive to this parameter. 
%Anyway, though microstructure study 
%showed that unresolved time structure could be as short as 10ns (Hankins 
%1996), which means that the radiation unit is very small, the scale of 
%the entire radiation region could be much larger. 
If $l_6$ is finally
determined to be much less than 1, then $P_{\rm sing,c}$ should be raised.
Note that the typical value of $F_\nu$(mJy) $d$(kpc) is larger than 
1. This also raises $P_{\rm sing,c}$. However, two effects, namely the
beaming effect $(4\pi f)$ and the enhancement of another factor $\xi$ 
could compensate the enhancement of $P_{\rm sing,c}$. Thus $\sim 10^{-4}
-10^{-3} {\rm erg}\cdot {\rm s}^{-1}$ could be a fair critical value of 
the emission power of the particles.

Such a critical value is too high for a less efficient
mechanism (e.g. curvature radiation, see next section), but is not
unmountable in principle. Using the constraint that the total emission 
energy over the emitting length is less than the kinetic energy of the 
particle, i.e., $P_{\rm sing,c}\cdot s/c <(\gamma-1)mc^2$, we obtain a 
very loose constraint as $\gamma-1 > P_{\rm sing,c}\cdot s/ mc^3$ 
$>5.7\times 10^{-3}$ $(4\pi f)F_\nu({\rm mJy}) d^2({\rm kpc})\Delta
\nu_9 PB_{12}^{-1} \xi^{-1}l_6^{-2}$. 

Next we'll examine the concrete direct emission mechanisms 
to see whether the high power defined by [6] is achievable.

\section{Curvature radiation (CR) and inverse Compton scattering (ICS)} 

The curvature radiation (CR) has long been regarded as a promising candidate 
to explain pulsar radio emission (e.g. Sturrock 1971, Ruderman \&  
Sutherland 1975), since its characteristic frequency 
$\nu_{\rm c,CR}=(3/ 4\pi)(\gamma^3 c/ \rho)\simeq 7.2\times10^{10} 
{\rm Hz}\gamma_3^3\rho_8^{-1}$ 
is just well within the observed emission band of pulsars.  
However, the emission power of a single particle 
$$P_{\rm sing,CR}={2\over 3}\gamma^4{e^2c\over \rho^2} 
\simeq 4.6\times 10^{-13}{\rm erg\cdot s}^{-1}\gamma_3^4\rho_8^{-2} \eqno(7)$$ 
is much lower than $P_{\rm sing,c}$ in eq.[6]. Thus incoherent CR is 
definitely ruled out as the candidate for pulsar radio emission (see also 
Lesch et al. 1998). This might be the
primary motivation of introducing ``bunching'' coherent mechanisms in the
early studies. It is worth noting here that even the coherent CR mechanism 
has been criticized from the observational points of views (Lesch et al. 1998).
Another objection is, when taking into account the energy loss of the 
secondaries, the required Lorentz factor of particles in the CR mechanism 
may lie above the ``Lorentz platform'', which makes it impossible to explain 
the low frequency emission from some pulsars (Zhang et al. 1997b).

Another direct emission model is the inverse Compton scattering model proposed
by Qiao (1988) and QL98. In such a model, a low frequency electro-magnetic 
wave is assumed to exist in the pulsar magnetosphere near the polar cap 
region, which is generated either by the breaking down of the vacuum polar 
gap (Ruderman \& Sutherland 1975; Zhang et al. 1997a,b) or other sorts of 
short-time oscillations or micro-instabilities (Bj\"ornsson 1996). Such a
wave is assumed to be inverse Compton scattered to the radio frequency we
observe. The typical emission frequency for the ICS process reads 
$\nu_{\rm c,ICS}=2\gamma^2\nu_{\rm gap}(1-\beta\cos\theta_i)$, 
where $\nu_{\rm gap}\sim c/h$ ($h$ is the gap height, typically $10^3$cm), 
$\theta_i$ is the incident angle of the scattering,  
and $\cos\theta_i=$ $[{2\cos\theta +(R/r)(1-3\cos^2\theta)}]/$  
$\{(1+3\cos^2\theta)[1-2(R/r)\cos\theta +(R/r)^2]\}^{1/2}$ (QL98). 
Near the polar cap region, the typical value for $(1-\beta\cos\theta_i)$ 
is $\sim 0.005$, thus the typical frequency of the ICS mechanism is
$\nu_{\rm c,ICS}\simeq 3.0\times10^9{\rm Hz}\gamma_2^2 h_3^{-1} 
{({1-\beta\cos\theta_i\over 0.005})}$, 
which is the typical frequency of pulsar radio emission. 
The emission power of a single particle due to ICS is $P_{\rm sing,ICS}
\simeq\gamma^2\sigma c U_{\rm ph}$, where $\sigma$ is the magnetized ICS 
cross section, the two polarized components of which are 
$\sigma(1)=\sin^2\theta_i/[\gamma^2(1-\beta\cos\theta_i)]
\sigma_{\rm th}\simeq 2\gamma^{-2}\sigma_{\rm th}$ for the regime of 
interests and $\sigma(2)=0$, and $U_{\rm ph}={1\over 4\pi}({2\Omega B 
h\over c})^2$ is the energy density of the gap-breakdown-induced low 
frequency waves (see details in QL98). With the Thomson cross section 
$\sigma_{\rm th}=6.65\times10^{-25}$cm$^2$, we finally obtain 
$$P_{\rm sing,ICS}\simeq 5.6\times 10^{-2}{\rm erg\cdot s}^{-1}B_{12}^2 
P_{0.1}^{-2}h_3^2. \eqno(8)$$ 
This is higher than the critical value in eq.[6].
Note both in eq.[6] and [8], the typical $P$ is adopted as 0.1s, which 
is the case for the majority of the pulsars. If $P$ is typically adopted as 
1s, then eq.[8] is less than [6], which means that energy budget constraint
still rules out the incoherent ICS process. This is consistent with the 
conclusions in QL98 (their Appendix C). Nevertheless, we have shown that
the ICS mechanism can marginally meet the energy budget requirement. 
 
\section{Conclusions} 
In this {\em Letter}, we have re-investigated the concepts of coherent
emission and posed two constraints on any incoherent emission mechanism
in pulsar magnetospheres. The energy budget constraint [2b] could be 
released by introducing an efficient radiation mechanism (e.g. ICS).
If the optical-thin constraint [2a] is also released, certain coherent 
mechanisms such as ``bunching'' or ``maser amplification'', are not 
obligatory any more. This in principle brings the possibility to 
interpret the high luminosities of pulsar in terms of
incoherent emission mechanism.
Though lack of an executable mechanism to release the optical-thin 
constraint without invoking any coherent mechanism, we suppose that
there might be some possible way out
after presenting some criticisms on the present studies. If one
could find a regime where $\tau <1$ [2a] is satisfied without 
invoking maser mechanism ($\tau <0$, a coherent mechanism), then
coherence is no longer essential to account for pulsar radio emission
provided an efficient enough mechanism.

We are not insisting here that the emission is definitely incoherent though. 
Besides the motivations to release the constraints [2a,b], coherence 
is also helpful to explain some other features of observations. For example, 
certain extent of coherence is required to account for the polarization 
features in the ICS model (e.g. Liu, Qiao, \& Xu 1998). Nevertheless, 
if the emission power of a single particle is high enough (e.g. ICS 
mechanism), and if the pulsar magnetospheres are really transparent,
the general criticisms (M78; M92a,b) on the bunching coherent 
mechanisms are greatly weakened, since a more efficient mechanism
(e.g. ICS) only need a much weaker degree of coherence, so that an 
instant or local bunching mechanism has been already adequate to  
account for the high luminosities. Even if the optical-thin constraint
can not be released unless certain coherent mechanism is incorporated, 
a high efficiency mechanism is also helpful to meet the requirement of 
the ``highest brightness temperature'' limit discussed in M92a.

\medskip
 
We thank the anonymous referee for his important criticisms to our original
manuscript, and Alice Harding for her careful reading the paper and  
insightful comments. Useful discussions with A.G.Lyne, J.L.Han, R.X.Xu and 
J.F.Liu are also appreciated. BZ acknowledges the NRC research associateship 
award and supports from China Postdoctoral Science Foundation. GJQ acknowledges
supports from NNSF of China, the Climbing Project of China, and the Project 
Supported by Doctoral Program Foundation of Institution of Higher Education 
in China. 
 
%\eject 
 
\end{document}